\documentclass[Conference]{IEEEtran}
\usepackage{amsfonts}
\usepackage[dvips]{graphicx}
\usepackage{epsfig,latexsym}
\usepackage{float}
\usepackage{indentfirst}
\usepackage{amsmath}
\usepackage{amssymb}
\usepackage{times}
\usepackage{subfigure}
\usepackage{stfloats,algorithm,algorithmic,bm}
\usepackage{psfrag}
\usepackage{cite}
\usepackage{subfigure}
\usepackage{stfloats}
\usepackage{bm}
\usepackage{colortbl}
\usepackage{xcolor}
\usepackage[section]{placeins}
\usepackage{caption}
\usepackage{cases}

\begin{document}

\title{A Diplexer-Based Receiver for Simultaneous Wireless Information and Power Transfer }

\author{\IEEEauthorblockN{Chong Qin, Yi Gong, and Zhi Quan}\\
\IEEEauthorblockA{Department of Electrical and Electronic Engineering\\
South University of Science and Technology of China\\
 qinc@mail.sustc.edu.cn, gongy@sustc.edu.cn, quanz@sustc.edu.cn}}

\maketitle

\begin{abstract}
\boldmath
Simultaneous wireless information and power transfer (SWIPT) is an appealing research area because both information and energy can be delivered to wireless devices simultaneously. In this paper, we propose a diplexer-based receiver architecture that can utilizes both the doubling frequency and baseband signals after the mixer. The baseband signals are used for information decoding and the doubling frequency signals are converted to direct current for energy harvesting. We analyze the signal in the receiver and find that the power of the energy harvested is equal to that of information decoded. Therefore, the diplexer can be used as a power splitter with a power splitting factor of 0.5. Specifically, we consider a multiuser multi-input single-output (MISO) system, in which each user is equipped with the newly proposed receiver. The problem is formulated as an optimization problem that minimizes the total transmitted power subject to some constraints on each user's quality of service and energy harvesting demand. We show that the problem thus formulated is a non-convex quadratically constrained quadratic program (QCQP), which can be solved by semi-definite relaxation.
%Numerical examples illustrate the impact of the quality of service and energy harvesting requests. It's observed that if the requests are higher, we need more transmitted power to satisfy these extra demands.
\end{abstract}


\section{Introduction}
Recently simultaneous wireless information and power transfer has become an attractive research area because it utilizes radio frequency (RF) signals both transferring information and delivering energy. The idea of SWIPT was first proposed in \cite{Varshney2008} and later the work is extended to frequency-selective channels with additive white Gaussian noise (AWGN) in \cite{GroverSahai2010}. The receiver is assumed to be able to decode information and harvest energy at the same time. However, the practical receiver circuit cannot meet the assumption as circuits for harvesting energy from RF signals cannot decode the carried information directly. Thus, \cite{ZhouZhangHo2013} and \cite{ZhangHo2013} proposed two practical schemes, namely power splitting (PS) and time switching (TS), respectively. The PS scheme divides the received signal into two parts according to an alterable PS factor, with one used for energy harvesting (EH) and the other one used for information decoding (ID). For the TS scheme, the receiver operates on the EH or ID mode at one time and the TS scheme can be considered a special case of the PS scheme with a PS factor of 0 or 1.

The traditional receiver removes the high frequency (HF) component after the mixer using low-pass filters, and reserves the baseband component for information decoding. In this paper, we propose a new receiver architecture equipped with a diplexer, which can utilize both the baseband component and the HF component simultaneously. In this newly proposed receiver, the HF component that should have been removed in traditional receivers can be transformed into direct current to charge the battery. The diplexer is used after the mixer in the RF chain to split the mixed doubling frequency  and baseband signals. In this way, both HF and baseband signal components can be utilized to achieve SWIPT.

With the diplexer-based receiver, we study a downlink multiuser MISO system that aims to minimize the transmitted power subject to the demands of ID and EH at the receivers. SWIPT has been recently studied in literature, i.e.,  \cite{ShiLiuXuZhang2014,TimotheouKrikidisZhengOttersten2014,KhandakerWong2014,ZhangLiHuangYang}. A multi-casting system was studied in \cite{KhandakerWong2014}, where the transmitter sends the same information to all users. In \cite{TimotheouKrikidisZhengOttersten2014}, the transmitter employs a linear precoding scheme to deliver information to different users, which use PS schemes and single user detection at the receivers. This paper optimizes both the beamforming and PS factors to minimize the total transmitted power subject to the individual quality-of-service (QoS) and EH thresholds. Coincidentally, \cite{ShiLiuXuZhang2014} studied the same problem but with more mathematical analysis. In addition, the security problem in SWIPT MISO systems was studied in \cite{ZhangLiHuangYang}, which aims to minimize the total transmit power subject to some constraints on the secrecy rate and energy harvested at each receiver.

The contributions of this paper are summarized as follows.

1) We propose a diplexer-based receiver architecture to achieve SWIPT. This receiver can utilize the HF component that should have been wasted previously. The PS factor of the receiver is a constant number 0.5.

2) A downlink multiuser MISO system with each user equipped with the diplexer-based receiver is investigated. The objective is to  minimize the total transmitted power subject to constraints on each user's quality of service and energy harvesting demand. The problem thus formulated is a quadratically constrained quadratic program (QCQP), and can be solved by SDP relaxation.

The rest of the paper is organized as follows. Section $ \text{\uppercase\expandafter{\romannumeral2}} $  introduces the receiver architecture and the system model. Section $ \text{\uppercase\expandafter{\romannumeral3}} $ formulates the problem of interest. Section $ \text{\uppercase\expandafter{\romannumeral4}} $ solves the problem of minimizing the total transmit power via SDP relaxation. Section $ \text{\uppercase\expandafter{\romannumeral5}} $ provides simulation results, and section $ \text{\uppercase\expandafter{\romannumeral6}} $ concludes the paper.
\section{System Model}

\subsection{Point to Point Wireless Link}%先介绍点对点无线通信系统，推出接收的信号
\begin{figure}[H]
  \begin{center}
  \includegraphics[height=1.5in,width=3.5in]{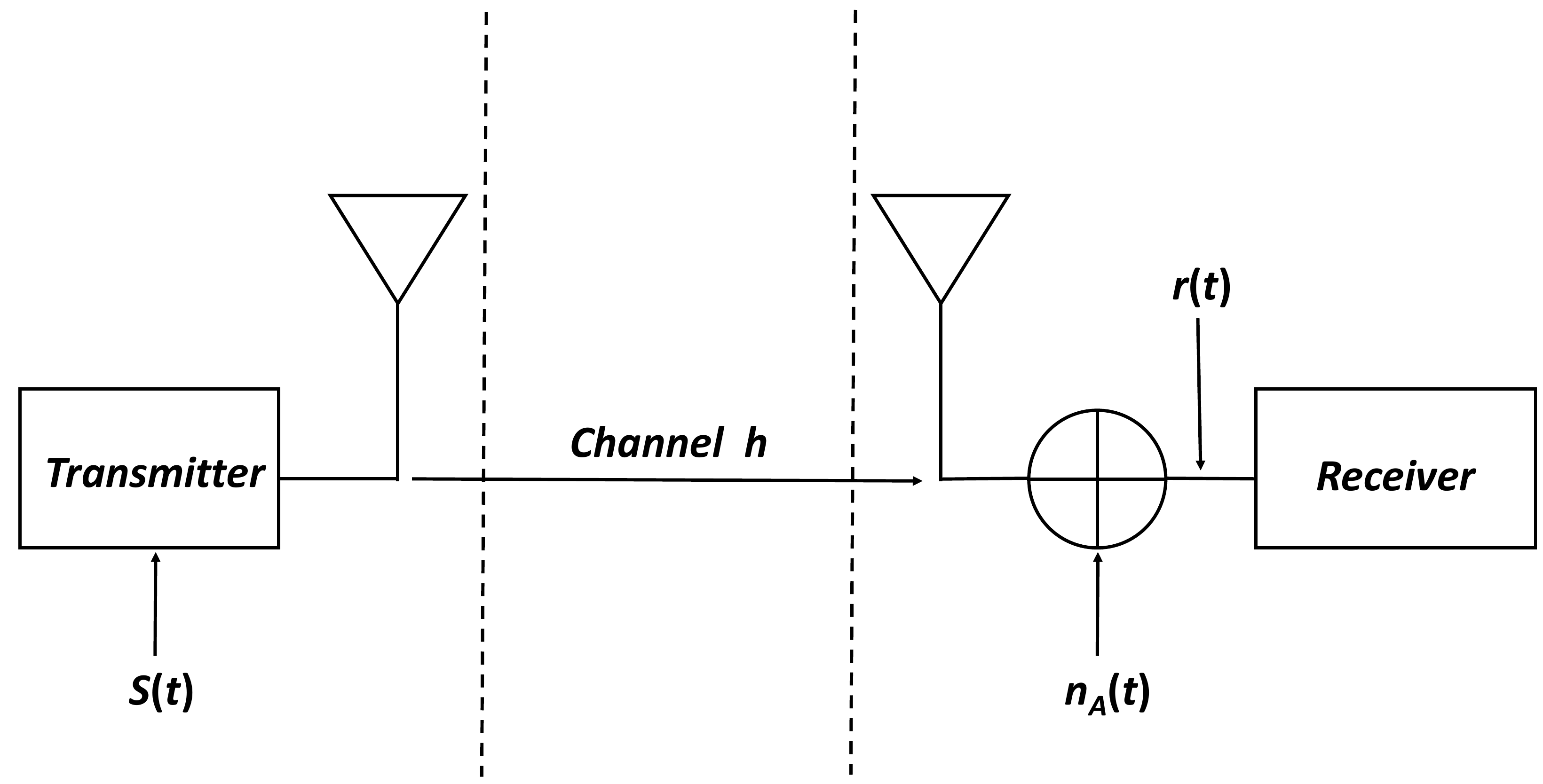}\\
  \caption{Point to point channel model.}
  \end{center}
\end{figure}
As  shown  in  Fig. 1,  we first study  a  point-to-point  wireless  communication  system  where  both  the  transmitter  and  the receiver are equipped with only one antenna. The received signal is denoted as $ r(t)$. The complex baseband message signal at the transmitter is denoted as $ m(t) $, whose amplitude and phase are real signals $ a(t) $ and $ b(t) $,
respectively, i.e., $ m(t)=a(t)e^{jb(t)}$. The average power of $ m(t) $ is normalized as $ \bf{\sl{E}}$$ \{ a^2 (t) \} = 1 $, where $ \bf{\sl{E}} $$ \{ . \}$ denotes mathematical expectation. Then, the transmitted RF bandpass signal $ S(t) $
 with carrier frequency
$ f_c $ is expressed as $ S(t)= \sqrt{2P_{avg}}\bf{\Re} $$ \{ m(t)e^{j2\pi f_c t} \} $, where $ \bf{\Re} \{ . \}$ and $ P_{avg} $
denote the real part of a complex number and the average transmit power of $ S(t) $, respectively.
The constant $ \sqrt{2} $ plays a role in keeping the power of $ S(t) $ consistent with
$ m(t) $. Without loss of generality, we assume that the bandwidth of $ S(t) $ is  $ B $ Hz, which is much lower than $ f_c $, i.e., $ B \ll f_c $.

%介绍信道和噪声
Assume a quasi-static flat fading channel  $ h=\sqrt{A} e^{j\phi} $,
where $ A(A>0) $ and $ \phi \in[0,2\pi) $  denote the channel power gain and phase shift, respectively.
The additive noise $ n_A(t) $ at the receiver is a bandpass narrow-band gaussian signal and is expressed as
$ n_A(t)=\sqrt{2}\bf{\Re} $$ \{ n_{Al}(t)e^{j2\pi f_c t} \} $, where $ n_{Al}(t)=n_I(t)+jn_Q{t} $ with $ n_I(t) $ and
$ n_Q(t) $ denote the in-phase and quadrature-phase components, respectively. $ n_I(t) $ and $ n_Q(t) $ are independent
identically distributed (i.i.d.) Gaussian random variables (RVs) with zero mean and variance $ \sigma_A^2/2 $.
Due to the independence of $ n_I(t) $ and $ n_Q(t) $, the power of $ n_{Al}(t) $ can be easily shown to be $ \sigma_A^2 $. Thus, $ n_{Al}(t) $ is a complex Gaussian RV with zero mean and variance $ \sigma_A^2 $, i.e., $ n_{Al}(t) \sim $ $ CN(0, \sigma_A^2 ) $.

%推出接收信号
From Appendix A, the received bandpass signal $ r(t) $ is given by
\begin{equation}\begin{aligned}
r(t)=&\sqrt{2AP_{avg}}a(t)cos(2\pi f_c t+b(t)+\phi)+\sqrt{2}c(t)cos(2\pi \\
&f_c t+d(t))
\end{aligned}\end{equation}
where real signals $ c(t) $ and $ d(t) $ denote the amplitude and phase of
$ n_{Al}(t) $, respectively, i.e., $ n_{Al}(t)=c(t)e^{jd(t)} $, $ \phi=-2\pi f_c \tau $ is the phase shift caused by the propagation delay.
%\setlength\abovedisplayskip{1pt}
%\setlength\belowdisplayskip{1pt}
%\vspace{-0.74cm}
\subsection{Receiver Architecture}
%引出自己设计的接收机架构
There are baseband and doubling frequency signal components at the mixer output of a coherent receiver. The traditional receiver removes the HF component after the mixer using low-pass filters, and reserves the baseband component for information decoding. In this paper, we propose a new receiver architecture equipped with a diplexer, which can utilize both the baseband component and the HF component simultaneously. In this proposed receiver, the HF component that should have been removed in traditional receivers can be transformed into direct current to charge the battery. The diplexer is used after the mixer in the RF chain to split the mixed doubling frequency  and baseband signals. In this way, both HF and baseband signal components can be utilized to achieve SWIPT.
Diplexer is a RF device used in many electronic products \cite{MichaelSteer2010}. It is a three-port network that splits incoming signals from a common port into two paths (baseband path and HF path) based on their frequency differences.

\begin{figure*}[htb]
  \begin{center}
  \includegraphics[height=2in,width=5in]{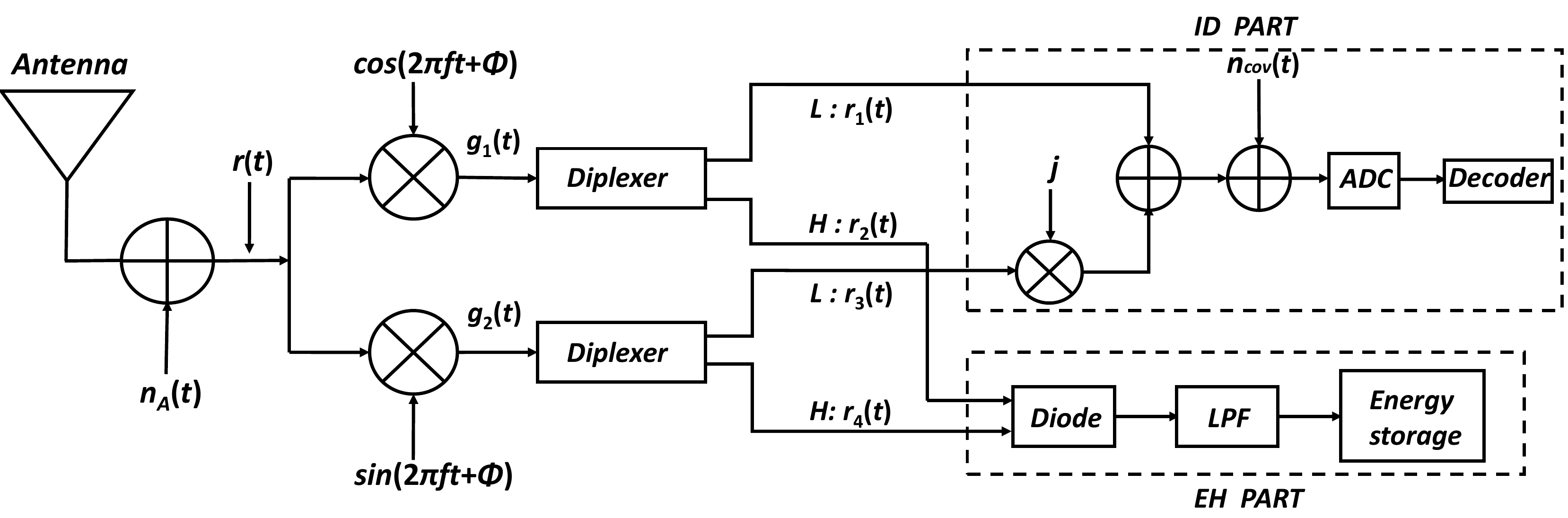}
  \caption{Receiver architecture with diplexers.}
  \end{center}
\end{figure*}

The proposed receiver architecture is shown in Fig. 2. Compared to a conventional ID receiver, we mainly replace the low-pass filters (LPF) with diplexers \cite{ZhouZhangHo2013}.
We assume that the receiver knows the phase shift $ \phi $ of the channel perfectly and it adopts coherent demodulation to demodulate the received signal. From $ r(t) $, local oscillation signals $ cos(2\pi f_c t+\phi) $ and $ sin(2\pi f_c t+\phi) $,  two real signals $ g_1(t) $ and $ g_2(t) $ are generated at the mixer output, and they are both composed of baseband and doubling frequency  components. Define $ g_1(t)\triangleq r_1(t)+r_2(t) $ and $ g_2(t)\triangleq r_3(t)+r_4(t) $
with $ r_1(t) $, $ r_3(t) $ denoting the corresponding baseband components and $ r_2(t) $, $ r_4(t) $ denoting the corresponding doubling frequency components, respectively. It is clear that $ r_1(t) $ and $ r_3(t) $ are the in-phase and quadrature components of $ r(t) $. We can recover $ r_l(t)=r_1(t)+j*r_3(t) $ for ID, where $ r_l(t) $ is the equivalent lowpass signal of $ r(t) $.
 The remaining HF signal component
$ H(t)\triangleq r_2(t)+r_4(t) $ is used to charge the energy storage for EH. It is worth noting that
this HF component is not used for EH in conventional SWIPT receivers.

Based on (1) and the standard trigonometric identities, $ g_1(t) $ and $ g_2(t) $ are rewritten as
\begin{equation}
\begin{split}
g_1(t)%=&r(t)*cos(2\pi f_c t+\phi)\\
      =&{\sqrt{\frac{AP_{avg}}{2}}}a(t)\{cos[4\pi f_c t+2\phi+b(t)]+cos[b(t)]\}\\
       &+\sqrt{\frac{1}{2}}c(t)\{cos[4\pi f_c t+\phi+d(t)]+cos[d(t)-
       \phi]\}\\
\end{split}
\end{equation}
\begin{equation}
\begin{split}
g_2(t)%=&r(t)*sin(2\pi f_c t+\phi)\\
      =&{\sqrt{\frac{AP_{avg}}{2}}}a(t)\{sin[4\pi f_c t+2\phi+b(t)]-sin[b(t)]\}\\
       &+\sqrt{\frac{1}{2}}c(t)\{sin[4\pi f_c t+\phi+d(t)]+sin[\phi-
       d(t)]\}.
\end{split}
\end{equation}

As a result, we have
\begin{equation}
r_1(t)={\sqrt{\frac{AP_{avg}}{2}}}a(t)cos[b(t)]+\sqrt{\frac{1}{2}}c(t)cos[d(t)-\phi]\\
\end{equation}
\begin{equation}
\begin{split}
r_2(t)=&{\sqrt{\frac{AP_{avg}}{2}}}a(t)cos[4\pi f_c t+2\phi+b(t)]+\sqrt{\frac{1}{2}}c(t)\\
&\{cos[4\pi f_c t+\phi+d(t)]\}\\
\end{split}
\end{equation}
\begin{equation}
r_3(t)={-\sqrt{\frac{AP_{avg}}{2}}}a(t)sin[b(t)]-\sqrt{\frac{1}{2}}c(t)sin[d(t)-\phi]\\
\end{equation}
\begin{equation}
\begin{split}
r_4(t)=&{\sqrt{\frac{AP_{avg}}{2}}}a(t)sin[4\pi f_c t+2\phi+b(t)]+\sqrt{\frac{1}{2}}c(t)\\
&{sin[4\pi f_c t+\phi+d(t)]}.
\end{split}
\end{equation}

For notational convenience, let $ L(t)\triangleq r_1(t)+r_3(t) $ and $ H(t) $ denote the
 baseband and HF components of $ G(t) $, respectively, where $ G(t) $ is defined as $ G(t)\triangleq g_1(t)+g_2(t) $. Let
$ \theta_1\triangleq b(t), \theta_2\triangleq d(t)-\phi, \theta_3\triangleq 4\pi f_c t+2\phi+b(t), $ and $  \theta_4\triangleq 4\pi f_c t+\phi+d(t). $
From Appendix B, we have
\begin{equation}\begin{aligned}
&L(t)=-\sqrt{AP_{avg}}a(t)sin(\theta_1-\pi/4)-c(t)sin(\theta_2-\pi/4)\\
&H(t)=\sqrt{AP_{avg}}a(t)sin(\theta_3+\pi/4)+c(t)sin(\theta_4+\pi/4)
\end{aligned}\end{equation}
% 平均功率之比 $ \bf{\sl{E}}$$ \{ H(t)^2 \}$/$ \bf{\sl{E}}$$ \{ L(t)^2 \}$.

From Appendix C, the average powers of $ L(t) $ and $ H(t) $ are found as
\begin{equation}\begin{aligned}
  {\bf{\sl{E}}}  \{ L^2(t)\}={\bf{\sl{E}}}  \{ H^2(t)\}  &= \frac{AP_{avg}}{2}+\frac{\sigma_A^2}{2}
\end{aligned}\end{equation}
which indicates that with the proposed diplexer-based receiver, the powers used for EH and for ID are the same. Let $ \rho $ denote the power splitting factor for EH and it is given by
\begin{equation}\begin{aligned}
\rho &=\frac{{\bf{\sl{E}}}\{H^2(t)\}} {{\bf{\sl{E}}}\{L^2(t)\}+{\bf{\sl{E}}}\{H^2(t)\}}=0.5.
\end{aligned}\end{equation}
%\vspace{-7pt}

{\textbf{Remark:}} The newly proposed receiver is able to utilize both the baseband component and the HF component simultaneously. In the proposed receiver, the HF component that should have been removed in traditional receivers can be transformed into direct current to charge the battery for EH.

As a result, the rectifier circuit can make full use of the power of $ H(t) $ for EH. Letting $\eta$ (0 $ <\eta\leq $ 1) denote the rectifier coefficient and ignoring the power of additive noise, the average power of harvested energy $Q$ is given by
\begin{equation}
Q=\frac{\eta AP_{avg}}{2}.
\end{equation}

By assuming that the conversion noise from passband to baseband in the ID process is $ n_{cov}(t)\sim CN(0,\sigma_{cov}^2) $, the average SNR is thus given by
\begin{equation}\begin{aligned}
SNR &=\frac{AP_{avg}}{\sigma_A^2+2\sigma_{cov}^2}=\frac{AP_{avg}}{\sigma^2} \\
\end{aligned}\end{equation}
where the equivalent ID noise is defined as $ \sigma^2\triangleq\sigma_A^2+2\sigma_{cov}^2 $.
\begin{figure}[H]
  \begin{center}
  \includegraphics[height=2in,width=2.5in]{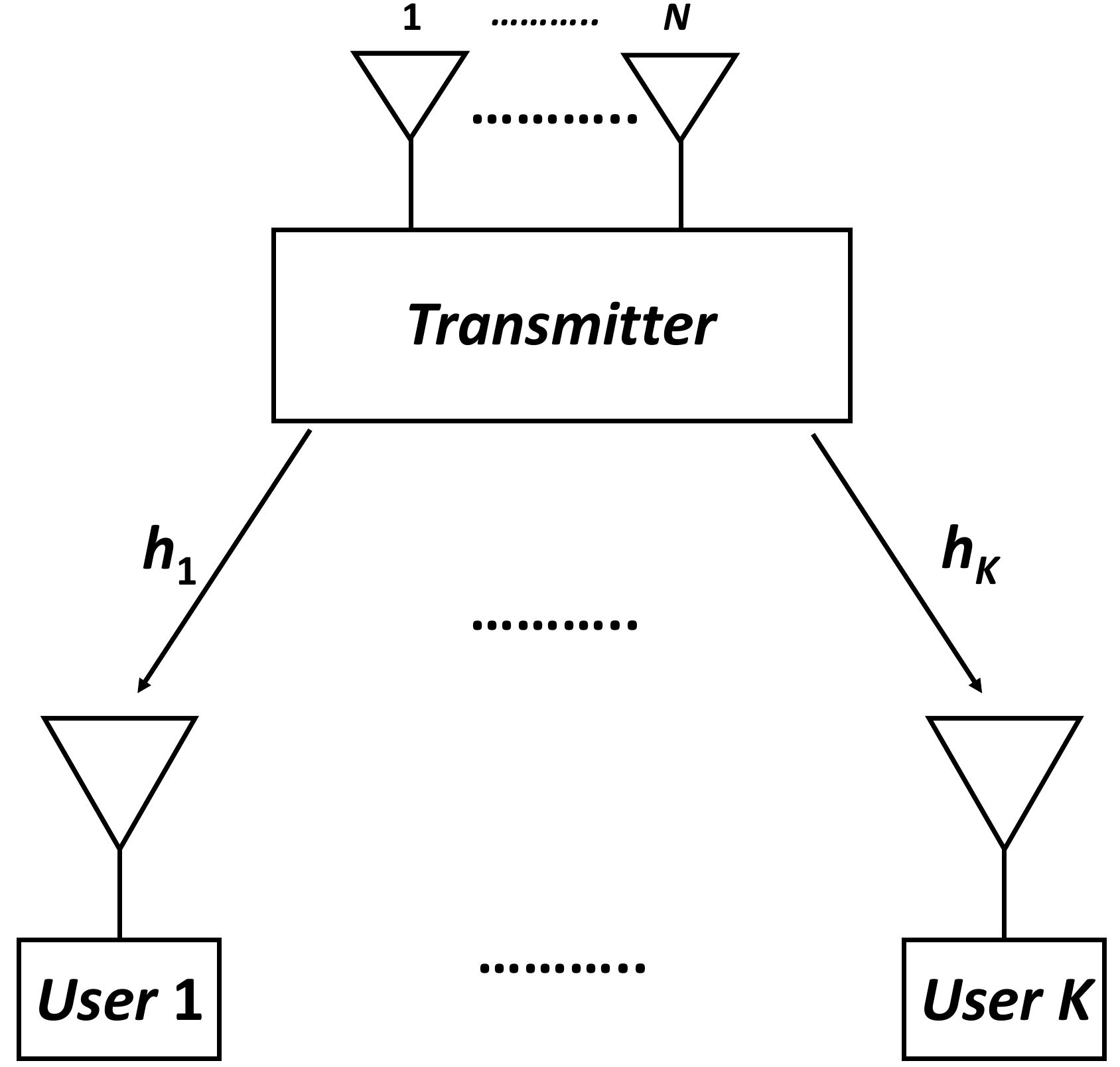}\\
  \caption{MISO channel model.}
  \end{center}
\end{figure}

\subsection{Multiuser MISO System}

Now, we apply the diplexer-based receiver into a downlink MISO system shown in Fig. 3, where there are one transmitter and multiple users. The transmitter is equipped with $ N $ $ (N > 1) $ antennas while $ K $ $ (K \geq 1) $ users each has one antenna.
The transmitter aims to transmit dedicated information to different users and the users employ the proposed receiver architecture to receive information and energy simultaneously. Without loss of generality, we assume that all the channels are subject to  frequency non-selective
block Rayleigh fading. $ {\bf{h}}_k $ $ (k\in\{1\ldots K\}) $ denotes the channel vector from the transmitter to the $ k_{th} $ user. With linear precoding at the transmitter, the received signal at the $ k_{th} $ user is given by
\begin{equation}
y_k={\bf{h}}_k^H\left(\sum_{m=1}^K{\bf{\omega}}_m s_m\right)+n_k
\end{equation}
where $ s_m\sim CN(0,1) $ is the transmitted data symbol to the $ m_{th} $ user (assumed to be independent for different users), $ {\bf{\omega}}_m $
is the beamforming coefficient corresponding to $ s_m $, $ n_k\sim CN(0,\sigma_A^2) $ is the additive noise at the $ k_{th} $ user.
With single user detection, the SINR at the $ k_{th} $ user is given by
\begin{equation}
SINR_k=\frac{|{\bf{h}}_k^H{\bf{\omega}}_k|^2}{\sum_{j\neq k}|{\bf{h}}_k^H{\bf{\omega}}_j|^2+\sigma^2}.
\end{equation}

On the other hand, the harvested energy at the $ k_{th} $receiver is given by
\begin{equation}
Q_k=\frac{1}{2}\eta\sum_{j=1}^{K}|{\bf{h}}_k^H{\bf{\omega}}_j|^2.
\end{equation}

\section{Optimization Problem}
 In this section, we consider an optimization problem that focuses on the beamformings design, aiming to minimize the total transmitted power subject to
 each user's EH and QoS constraints. The optimization problem {$ \bf{P0} $} is formulated as
%\\{\bf{P0}:}\qquad
\begin{equation}
\begin{array}{l}
\begin{array}{*{20}{c}}
{\mathop {\min }\limits_{\left\{{\bf{\omega_j}}\right\}_{j=1}^K} }&{\sum_{j=1}^K|{\bf{\omega_j}}|^2}
\end{array}\\
\begin{array}{*{20}{c}}
{\text{S.T.}}&\begin{array}{l}\\
SINR_k \geq \gamma_k \qquad\forall k=1,2,\ldots,K\\
Q_k \geq \mu_k \qquad\forall k=1,2,\ldots,K\\
\end{array}&{}
\end{array}
\end{array}
\end{equation}
where $ SINR_k $ and $ Q_k $ are given in (14) and (15). $ \gamma_k $ and $ \mu_k $ denote the SINR and EH threshold of the $ k_{th} $ user, respectively.
\subsection{Feasibility Test}
We now check the feasibility of $ {\bf{P0}} $. Once the SINR constraint is satisfied, we can always find an amplification factor and multiply it with the beamforming vector to satisfy the EH constraint. Therefore, if the SINR constraint is met, $ {\bf{P0}} $ will be feasible. It follows from \cite{HungerJoham2010} that if and only if the SINR thresholds satisfying $ \sum_{k=1}^K\frac{\gamma_k}{\gamma_k+1}\leq rank(\bf{H}) $, $ {\bf{P0}} $ with only SINR constraints is feasible, where $ \bf{H} $ is defined as $ \bf{H}\triangleq[h_1\quad h_2 \ldots h_K] $. We assume that $ {\bf{P0}} $ is feasible in the remaining sections.
\subsection{Convexity analysis}
Next, we analyze the convexity of {$ \bf{P0} $}. The SINR constraint is convex since it can be converted to second-ordered cone restrictive conditions as shown in \cite{LuoYu2006}. The EH constraint is non-convex since it can be seen as the sum of concave functions. Therefore, $ \bf{P0} $ is non-convex and belongs to the class of non-convex QCQP problems. Without loss of generality, we use semidefinite programming (SDP) relaxtion method to approach the optimal solutions.

\section{SDP Relaxation}
We first introduce the new matrixes $ {\bf{W}}_j={\bf{\bf\omega}}_j{\bf{\omega}}_j^H $ and $ {\bf{H}}_j={\bf{h}}_j {\bf{h}}_j^H $. Recall that the inner product of two hermitian matrixes $ {\bf{A}} $ and $ {\bf{B}} $ is $ \textrm{Tr}({\bf{AB}}) $ and $ {\bf{x}}^H{\bf{Ax}}=\textrm{Tr}({\bf{Axx}}^H) $. Thus, we can convert $ {\bf{P0}} $ to $ {\bf{P1}} $:
%\\{\bf{P1}:}\qquad %问题P1
\begin{equation}
\begin{array}{l}
\begin{array}{*{20}{c}}
{\mathop {\min }\limits_{\left\{{\bf{W}}_j\right\}_{j=1}^K} }&{\sum_{j=1}^K \textrm{Tr}({\bf{W}}_j)}
\end{array}\\
\begin{array}{*{20}{c}}
{\text{s.t.}}&\begin{array}{l}\\
\textrm{Tr}({\bf{H}}_k {\bf{W}}_k)-\gamma_k\sum_{j\neq k}\textrm{Tr}({\bf{H}}_k {\bf{W}}_j)\geq\gamma_k\sigma^2  \quad\forall k \\\\
\sum_{j=1}^K \textrm{Tr}({\bf{H}}_k {\bf{W}}_j)\geq \xi_k \quad\forall k,\\\\
{\bf{W}}_k \succeq 0 \quad\forall k \\\\
rank({\bf{W}}_k)=1 \quad\forall k,
\end{array}&{}
\end{array}
\end{array}
\end{equation}
where $ \xi_k $ is defined as $ \xi_k\triangleq 2\mu_k/\eta $ and can be regarded as a new EH threshold. Dropping the rank constraint, we have $ {\bf{P2}} $:
%\\{\bf{P2}:}\qquad
\begin{equation}
\begin{array}{l}
\begin{array}{*{20}{c}}
{\mathop {\min }\limits_{\left\{{\bf{W}}_j\right\}_{j=1}^K} }&{\sum_{j=1}^K \textrm{Tr}({\bf{W}}_j)}
\end{array}\\
\begin{array}{*{20}{c}}
{\text{s.t.}}&\begin{array}{l}\\
\textrm{Tr}({\bf{H}}_k {\bf{W}}_k)-\gamma_k\sum_{j\neq k}\textrm{Tr}({\bf{H}}_k {\bf{W}}_j)\geq\gamma_k\sigma^2  \quad\forall k \\\\
\sum_{j=1}^K \textrm{Tr}({\bf{H}}_k {\bf{W}}_j)\geq \xi_k \quad\forall k \\\\
{\bf{W}}_k \succeq 0 \quad\forall k,\\\\
%rank({\bf{W_k}})=1 \quad\forall k,
\end{array}&{}
\end{array}
\end{array}
\end{equation}
Let
\begin{equation}
\begin{split}
&\beta_k=\gamma_k\sigma^2 \\
&{\bf{G}}_k=-\gamma_k {\bf{H}}_k \\
&{\bf{M}}_j= \left\{ {\begin{array}{*{20}{c}}
{\bf{H}}_j\quad j=k \\
{\bf{G}}_j\quad j\neq k
\end{array}}\right..
\end{split}
\end{equation}\\
{\bf{P2}} can be transformed to a concise form {\bf{P3}}:
%\\{\bf{P3}:}
\begin{equation}
\begin{array}{l}
\begin{array}{*{20}{c}}
{\mathop {\min }\limits_{\left\{{\bf{W}}_j\right\}_{j=1}^K} }&{\sum_{j=1}^K \textrm{Tr}({\bf{W}}_j)}
\end{array}\\
\begin{array}{*{20}{c}}
{\text{s.t.}}&\begin{array}{l}\\
\sum_{j=1}^K \textrm{Tr}({\bf{M}}_j {\bf{W}}_j)\geq \beta_k \quad\forall k\\\\
\sum_{j=1}^K \textrm{Tr}({\bf{H}}_k {\bf{W}}_j)\geq \xi_k \quad\forall k\\\\
{\bf{W}}_k \succeq 0 \quad\forall k,\\
\end{array}&{}
\end{array}
\end{array}
\end{equation}
which can be further written as {\bf{P4}:}
%\\{\bf{P4}:}\qquad\\ %问题P4
\begin{equation}
 \begin{array}{l}
\begin{array}{*{20}{c}}
{\mathop {\min }\limits_{\left\{{\bf{W}}_j\right\}_{j=1}^K} }&{\sum_{j=1}^K \textrm{Tr}({\bf{W}}_j)}
\end{array}\\
\begin{array}{*{20}{c}}
{\text{s.t.}}&\begin{array}{l}\\%下面是加限制条件的俄地方

{\bf{A}}_k=\left[ {\begin{array}{*{20}{c}}
{\sum_{j=1}^K\textrm{Tr}({\bf{M}}_j {\bf{W}}_j)} & \sqrt{\beta_k}\\\\
\sqrt{\beta_k} & 1
\end{array}} \right] \succeq 0 \quad\forall k \\\\

{\bf{B}}_k=\left[ {\begin{array}{*{20}{c}}
{\sum_{j=1}^K \textrm{Tr}({\bf{H}}_k {\bf{W}}_j)} & \sqrt{\xi_k}\\\\
\sqrt{\xi_k} & 1
\end{array}} \right] \succeq 0 \quad\forall k \\\\
{\bf{W}}_k \succeq 0 \quad\forall k.
\end{array}&{}
\end{array}
\end{array}
\end{equation}

$ {\bf{P4}} $ is a standard SDP problem can be solved using a SDP solver (i.e., CVX \cite{GrantBoyd2015}). If $ \left\{ {\bf{W}}_j \right\} $ are rank-one, then the optimal solutions of the original problem can be derived directly from them, i.e. eigenvalue decomposition (EVD).
%In our simulations, the relaxed problem always return rank-one solutions and we can obtain the original optimal solutions from them.
%\setlength\abovedisplayskip{1pt}
%\setlength\belowdisplayskip{1pt}
\section{Simulation Results}
In this section, we provide numerical results to evaluate the transmit beamforming scheme. The simulation scenario assumes that the transmitter is equipped with four antennas,
each of which is used for a user, i.e., $K=N=4$. All channels are Rayleigh fading and their pathlosses are assumed to be $-40$ dB. All users have the same thresholds on the QoS and energy harvested. The additive noise $ \sigma_A^2 $ is $-70$ dBm and the conversion noise $ \sigma_{cov}^2 $ is $-50$ dBm.
The EH conversion is assumed perfect.
\begin{figure}[!t]
  \begin{center}
  \includegraphics[height=3in,width=3.3in]{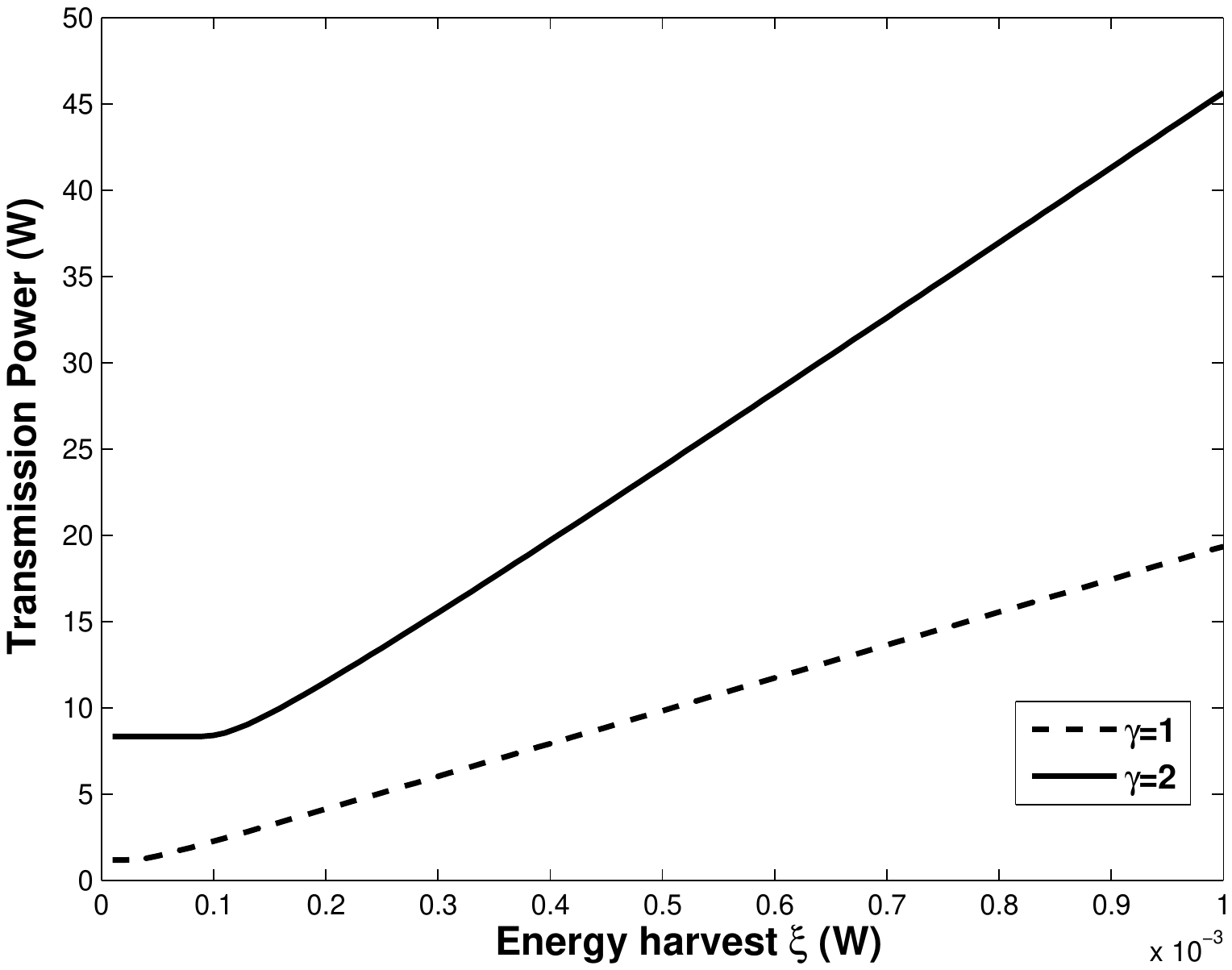}
  \caption{Transmission power VS EH threshold $ \xi $ }
  \end{center}
\end{figure}

In Fig. 4, we investigate the minimum transmission power versus the EH threshold $ \xi $ with the SINR threshold $ \gamma $ being fixed. For $ \gamma =2$, we can see that the transmitted power does not change as $\xi$ increases in the former part of the curve. This is because when the value of $\xi $ is small, the SINR constraint is tighter than the EH constraint and the SINR constraint mainly determines the transmitted power. When $\xi $ becomes larger, we need more transmission power to cover the EH demand. For $\gamma=1$, the SINR request is low and the EH constraint dominates the amount of the transmitted power. In Fig. 5, we study how the SINR threshold affect on the minimum transmission power with the EH threshold $ \xi $ being fixed. It can be seen that more energy is needed to satisfy the tighter SINR constraints.

\begin{figure}[!t]
  \begin{center}
 \includegraphics[height=3in,width=3.3in]{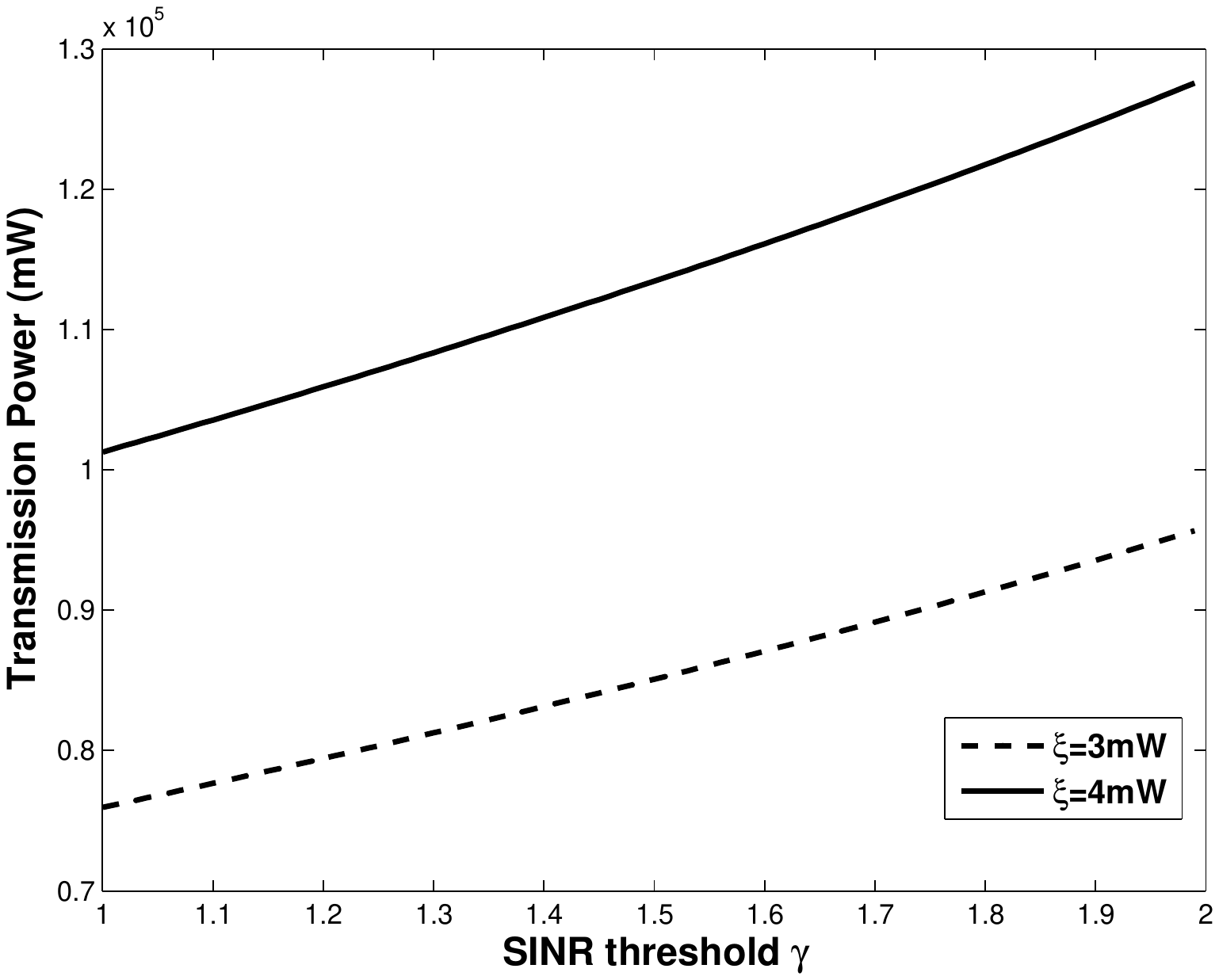}
  \caption{Transmission power VS SINR threshold $ \gamma $ }
  \end{center}
\end{figure}
%\vspace{-1.74cm}
\section{Conclusion}
%We propose an innovation receiver architecture with diplexers in the SWIPT field. Conventional receivers remove HF signals directly after mixers and the HF part just wastes without utilizing in any meaningful way. To tackle this problem, we propose a new receiver architecture added diplexers behind mixers to separate the doubling frequency and  baseband signals, and then use the HF and BB signals for EH and ID respectively. Thus the HF signals ignored previously are taken full advantage. Besides, the new receiver architecture is modified little based on conventional one and it's easy to implement it in practice. After investigating the signal flow in the receiver, we derive that the PS factor equals to 0.5 exactly. We employ the new receiver in MISO systems for SWIPT and aim to save the transmitted power subject to individual QoS and EH request. We use SDR method solving it and the simulation result always return rank one solutions, which means the SDR method maybe always tight. Finally, the numerical results show the influences of the EH and ID threshold to the total transmitted power.
We have proposed a diplexer-based receiver to utilize the HF signal component that should have been wasted previously. The new receiver architecture has a PS factor of 0.5. We deploy the new receiver into a MISO system, which aims to minimize the transmitted power subject to constraints on individual QoS and EH demands. The non-convex problem thus formulated can be solved using SDP relaxation.

\begin{appendices}
%\vspace{-1.04cm}
\section{}%Appendix A
%\vspace{-0.1cm}
Let $ \tau $ denote the propagation delay and assume $ \tau \ll 1/B $ for narrow-band signals. Let $ c(t) $ and $ d(t) $ denote the amplitude and phase of $ n_{Al}(t) $, respectively, i.e., $ n_{Al}(t)=c(t)e^{j d(t)} $. The corresponding phase shift is given by $ \phi=-2\pi f_c \tau $. It thus follows that
\begin{subequations}
\begin{align}
r(t)&\nonumber\\
=&\sqrt{A}S(t-\tau)+n_A(t) \\
=&\sqrt{2}{\bf{\Re}} \left\{\sqrt{AP_{avg}}m(t-\tau)e^{j2\pi f_c (t-\tau)}+n_{Al}(t)e^{j2\pi f_c t}\right\}\\
=&\sqrt{2}{\bf{\Re}} \left\{\sqrt{AP_{avg}}m(t)e^{j(2\pi f_c t+\phi)}+n_{Al}(t)e^{j2\pi f_c t} \right\}\\
=&\sqrt{2}{\bf{\Re}} \left\{\sqrt{AP_{avg}}a(t)e^{j[2\pi f_c t+\phi+b(t)]}+c(t)e^{j[j2\pi f_c t+d(t)]} \right\}\\
=&\sqrt{2AP_{avg}}a(t)cos[2\pi f_c t+b(t)+\phi]+\sqrt{2}c(t)cos[2\pi f_c t \\
 &+d(t)].\nonumber
\end{align}
\end{subequations}

We get $ 22(a) $ from the input-output channel model of continuous-time passband signals. By substituting the equivalent lowpass expressions of $ S(t) $ and $ n_{Al}(t) $, we obtain $ 22(b) $. After ignoring $ \tau $ and utilizing $ m(t)=a(t)e^{jb(t)} $, $ n_{Al}(t)=c(t)e^{jd(t)} $, and thus $ 22(d) $ is derived.
%\vspace{-0.74cm}
\section{}
\vspace{-0.34cm}
\begin{subequations}
\begin{align}
L(t)=&r_1(t)+r_3(t)\\
    =&\sqrt{AP_{avg}}a(t)(sin\frac{\pi}{4}cos\theta_1-sin\theta_1cos\frac{\pi}{4})+c(t)(sin\frac{\pi}{4}\nonumber\\
    &cos\theta_2-sin\theta_2cos\frac{\pi}{4})  \\ \nonumber
    =&-\sqrt{AP_{avg}}a(t)sin(\theta_1-\pi/4)-c(t)sin(\theta_2-\pi/4).
\end{align}
\end{subequations}%\nonumber

We get $ 23(a) $ through definition. After substituting the expressions of $ r_1(t) $ and $ r_3(t) $, $ 23(b) $ is obtained. Based on the standard trigonometric identities, we obtain $ 23(c) $. Similarly, we can obtain the expression of $ H(t) $ as follows
\begin{equation}
\begin{split}
H(t)=&r_2(t)+r_4(t)\\
    =&\sqrt{AP_{avg}}a(t)(sin\frac{\pi}{4}cos\theta_3+sin\theta_3cos\frac{\pi}{4})+c(t)(sin\frac{\pi}{4}\\
    &cos\theta_4+sin\theta_4cos\frac{\pi}{4})\\
    =&\sqrt{AP_{avg}}a(t)sin(\theta_3+\pi/4)+c(t)sin(\theta_2+\pi/4).
\end{split}
\end{equation}
%\vspace{-0.74cm}
\section{}
%\vspace{-0.34cm}
The average power of $ H(t) $ is given by
\begin{equation}
\begin{split}
&\quad{\bf{\sl{E}}}\{ H^2(t)\}\\
&= {\bf{\sl{E}}}  \{ {(\sqrt{AP_{avg}}}a(t)sin(\theta_3+\pi/4)+c(t)sin(\theta_4+\pi/4))^2 \} \\
&= {\bf{\sl{E}}}  \{{AP_{avg}} a^2(t)sin^2(\theta_3+\pi/4)+c^2(t)sin^2(\theta_4+\pi/4) \\
&\qquad+2\sqrt{AP_{avg}}a(t)c(t)sin(\theta_3+\pi/4)sin(\theta_4+\pi/4)\} \\
&= \frac{AP_{avg}}{2}+\frac{\sigma_A^2}{2}.
\end{split}
\end{equation}

The last equality holds because different random variables are independent and $ {\bf{\sl{E}}} \{sin(\theta_3+\pi/4)\}=0 $ when $ \theta_3 $ is uniformly distributed in $ [0,2\pi) $.

Similarly, we can calculate the average power of $ L(t) $
\begin{equation}
\begin{split}
&\quad{\bf{\sl{E}}}\{ L^2(t)\}\\
&= {\bf{\sl{E}}}  \{ {[\sqrt{AP_{avg}}}a(t)sin(\theta_1-\pi/4)+c(t)sin(\theta_2-\pi/4)]^2 \} \\
&= {\bf{\sl{E}}}  \{{AP_{avg}} a^2(t)sin^2(\theta_1-\pi/4)+c^2(t)sin^2(\theta_2-\pi/4) \\
&\qquad+2\sqrt{AP_{avg}}a(t)c(t)sin(\theta_1-\pi/4)sin(\theta_2-\pi/4)\} \\
&= \frac{AP_{avg}}{2}+\frac{\sigma_A^2}{2}.
\end{split}
\end{equation}

\end{appendices}
%论文引用


\begin{thebibliography}{99}

\bibitem{Varshney2008}
L. R. Varshney, \textquotedblleft Transporting information and energy simultaneously,\textquotedblright \space \emph{Proc. IEEE Int. Symp. Inf. Theory}, pp. 1612-1616, July 2008.

\bibitem{GroverSahai2010}
P. Grover and A. Sahai, \textquotedblleft Shannon meets Tesla: wireless information and power transfer,\textquotedblright \space \emph{Proc. IEEE 2010 Int. Symp. Inf. Theory}, pp. 2363-2367.

\bibitem{ZhouZhangHo2013}
X. Zhou, R. Zhang, and C. K. Ho, \textquotedblleft Wireless information and power transfer: architecture design and rate-energy tradeoff,\textquotedblright \space \emph{IEEE Trans. Commun.}, vol. 61, no. 11, pp. 4754-4767, Nov 2013.

\bibitem{ZhangHo2013}
R. Zhang and C. K. Ho, \textquotedblleft MIMO broadcasting for simutaneous wireless information and wireless transfer, \textquotedblright \space \emph{IEEE Trans. Wireless Commun.}, vol. 12, no. 5, pp. 1989-2001, May 2013.

\bibitem{ShiLiuXuZhang2014}
Q. J. Shi, L. Liu, W. Xu and R. Zhang, \textquotedblleft Joint transmit beamforming and receive power splitting for MISO SWIPT systems,\textquotedblright \space \emph{IEEE Trans. Wireless Commun.}, vol. 13, no. 6, pp. 3269-3280, June 2014.

\bibitem{HungerJoham2010}
R. Hunger and M. Joham, \textquotedblleft A complete description of the QoS feasibility region in the vector broadcast channel,\textquotedblright \space \emph{IEEE Trans. Signal Process.}, vol. 58, no. 7, pp. 3870-3878, July 2010.

\bibitem{TimotheouKrikidisZhengOttersten2014}
S. Timotheou, I. Krikidis, G. Zheng, and B. Ottersten, \textquotedblleft Beamforming for MISO Interference Channels with Qos and RF Energy Transfer,\textquotedblright \space \emph{IEEE Trans. Wireless Commun.}, vol. 13, no. 5, pp. 2646-2658, May 2014.

\bibitem{KhandakerWong2014}
M. R. A. Khandaker and K. K. Wong, \textquotedblleft SWIPT in MISO Multicasting Systems,\textquotedblright \space \emph{IEEE Wireless Commun Lett.}, vol. 3, no. 3, pp. 277-280, June 2014.

\bibitem{LuoYu2006}
Z. Q. Luo and W. Yu, \textquotedblleft An introduction to convex optimization for communications and signal processing,\textquotedblright \space \emph{IEEE J. Sel. Areas Commun.}, vol. 24, no. 8, pp. 1426-1438, August 2006.

\bibitem{ZhangLiHuangYang}
H. Y. Zhang, C. G. Li, Y. M. Huang, and L. X. Yang, \textquotedblleft Secure Beamforming for SWIPT in Multiuser MISO Broadcast Channel With Confidential Messages,\textquotedblright \space \emph{IEEE Commun Lett.}, vol. 19, no. 8, pp. 1347-1350, August 2015.

\bibitem{Goldsmith2005}
A. Goldsmith, \emph{Wireless Communications}, Cambridge University Press, 2005.

\bibitem{TseViswanath2005}
D. Tse and P. Viswanath, \emph{Fundamentals of Wireless Communication}, Cambridge University Press, 2005.

\bibitem{PalomarEldar2010}
D. P. Palomar and Y. Eldar, \emph{Convex Optimization in Signa Processing and Communications}, Cambridge University Press, 2010.

\bibitem{GrantBoyd2015}
M. Grant and S. Boyd, CVX: Matlab software for disciplined convex programming, version 3.0 beta, August 2015. Available: http://cvxr.com/cvx.

\bibitem{MichaelSteer2010}
M. Steer, \emph{Microwave and  RF Design}, SciTech Publishing, 2010.

\end{thebibliography}
\end{document}